\documentclass[11pt]{article}
\usepackage{amsmath,amssymb,epsfig,sint}
\font\sixrm=cmr6

\newcommand{\R}{\rm I\kern-.2emR}
\newcommand{\C}{\rm \kern.25em\vrule height1.4ex
depth-.12ex width.06em\kern-.31em C}
\newcommand{\N}{{\rm I\kern-.16em N}}
\newcommand{\Z}{{\rm Z\kern-.35em Z}}

\newcommand{\kappac}{\kappa_{\rm c}}

\newcommand{\gr}{g_{{\hbox{\sixrm R}}}}
\newcommand{\alphar}{\alpha_{{\hbox{\sixrm R}}}}
\newcommand{\grthree}{g_{{\hbox{\sixrm R}}3}}

\newcommand{\mr}{m_{{\hbox{\sixrm R}}}}

\newcommand{\Zr}{Z_{{\hbox{\sixrm R}}}}

\newcommand{\mrnaive}{\mr^{\rm naive}}

\newcommand{\rmd}{{\rm d}}
\newcommand{\rmO}{{\rm O}}

\newcommand{\be}{\begin{equation}}   
\newcommand{\ex}{\end{equation}}
\newcommand{\ba}{\begin{eqnarray}}
\newcommand{\ea}{\end{eqnarray}}

\newcommand{\Zrhat}{\widehat{\Zr}}
\newcommand{\grhat}{\widehat{\gr}}
\newcommand{\lambdahat}{{\widehat{\lambda}}}

\newcounter{subequation}[equation]
\makeatletter

\expandafter\let\expandafter
\reset@font\csname reset@font\endcsname

\def\subeqnarray{\arraycolsep1pt
    \def\@eqnnum\stepcounter##1{\stepcounter{subequation}%
        {\reset@font\rm(\theequation\alph{subequation})}}
\jot5mm     \eqnarray}

\makeatother

\newcommand{\msbar}{{\rm \overline{MS\kern-0.14em}\kern0.14em}}

\begin{document}
\begin{titlepage}

\begin{flushright}
   MPP-2006-1\\
   January 2006
\end{flushright}

\vskip 0.20 true cm

\begin{center}
{\Large\bf 
Repairing Stevenson's step in the $4d$ Ising model} 
\end{center}
\vskip 1 true cm
\centerline{\large Janos Balog}
\vskip1ex
\centerline{Research Institute for Particle and Nuclear Physics}
\centerline{1525 Budapest 114, Pf. 49, Hungary}
\vskip 1 true cm
\centerline{\large Ferenc Niedermayer${}^*$}
\vskip1ex
\centerline{Institute for Theoretical Physics, University of Bern}
\centerline{CH-3012 Bern, Switzerland}
\vskip 1 true cm
\centerline{\large Peter Weisz}
\vskip1ex
\centerline{Max-Planck-Institut f\"ur Physik}
\centerline{F\"ohringer Ring 6, D-80805 M\"unchen, Germany}
\vskip 1 true cm
\centerline{\bf Abstract}
\vskip 1.0ex
In a recent paper Stevenson claimed that analysis
of the data on the wave function renormalization constant
near the critical point of the $4d$ Ising model
is not consistent with analytical expectations. 
Here we present data with improved statistics 
and show that the results are indeed consistent
with conventional wisdom once one takes into account 
the uncertainty of lattice artifacts in the analytical computations.

\vfill
\noindent{---------------}\\
\noindent\footnotesize{${}^*$On leave from E\"otv\"os University, 
HAS Research Group, Budapest, Hungary} 
\eject

\end{titlepage}

\section{Introduction}

One of the apparently simplest quantum field theories in four 
dimensions is the $\phi^4$ theory with $n$ components. 
Conventional wisdom (CW) holds that the theory is trivial for all $n\ge1$
\footnote{in the sense that if the model is defined non-perturbatively 
with an ultraviolet cutoff $\Lambda$, say via lattice regularization,
connected $r-$point functions with $r>2$ vanish in the limit
$\Lambda\to\infty$}. Unfortunately there is presently no rigorous proof.
One relies heavily on the validity of renormalization group (RG) 
equations for physical quantities \cite{BGZ}
together with boundary conditions at finite cutoff 
provided by non-perturbative methods \cite{LWsymm,LWbroken}.

Apart from its purely theoretical interest it is 
of phenomenological relevance since for the case $n=4$ it constitutes
the pure Higgs sector of the Minimal Standard Model (MSM).
The fact that $\phi^4_4$ is (probably) trivial does not invalidate
(renormalized) perturbative computations for amplitudes 
at energies well below the physical cutoff where the MSM may be a good
effective theory. 

In the past triviality has been invoked to propose upper bounds on the
mass of the Higgs boson (see e.g. refs.~\cite{HJLNY-HKNV}). 
It must be stressed that these bounds are non-universal, 
they depend on the particular regularization.
But for a given regularization it is conventionally accepted 
that such a bound can be given a precise meaning.
In recent papers Cea, Cosmai, and Consoli (CCC) \cite{CCC,CCC2} 
claim that triviality itself cannot be used 
to place upper bounds on the Higgs mass
even for a given regularization. They assert that standard
predictions of the RG analysis for the behavior of some quantities
near the critical line are not valid. If true this would indeed be
rather important because it would reveal a serious flaw in
our conventional theoretical understanding of the pure $\phi^4_4$!

In ref.~\cite{BDNWW} Duncan, Willey and the present authors 
explained why critiques of the standard picture raised in 
ref.~\cite{CCC} were not relevant. Nevertheless one must admit 
that the unconventional picture of CCC cannot be ruled out by
present numerical simulations. Recently two papers appeared,
the first by CCC \cite{CCC2} and the second by Stevenson \cite{Stevenson}, 
again claiming finer but significant discrepancies between quantitative 
(standard) analytic predictions and numerical data in the $4d$ Ising 
model\footnote{The Ising model is obtained as the limit of the $\phi^4$
model when the bare coupling goes to infinity. As such one generally
considers this the ``worst case" i.e. if CW holds for the Ising
model it is even more plausible for finite bare coupling.}.  

It is the purpose of this paper to reply to these challenges 
and to demonstrate that they are too weak
to seriously cast doubt on CW.
The main objection by CCC and one by Stevenson are rather easy to dismiss.
A second objection by Stevenson is more difficult. 
It concerns a certain difference $\triangle$
between the wave function renormalization constants
below and above the critical point 
\be
\triangle=\Zrhat(\kappa=0.074)-\Zrhat(\kappa=0.0751)\,,
\end{equation}
which we have called ``Stevenson's step" in the title.
The $\kappa$ values chosen here are about the closest to the 
critical point where one can presently obtain good statistics with
some (but not unreasonable) computational effort
\footnote{the correlation lengths are around 6 in lattice units.}.
Stevenson claims that ``theoretical predictions for $\triangle$
cannot be pushed above $0.05$ well short of the ``experimental" 
value $\triangle_{\rm MC}=0.071(6)$". 
It is of course debatable whether such a small discrepancy 
indicates a potential problem, however we 
decided it merited more careful investigation. 

It is however clear that one is here addressing 
few percent effects, and to clarify the situation we need both data 
and analyses which are precise to this level. 
The main analytic sources of error concern
the treatment of higher order cutoff effects in the framework
of renormalized PT. The main sources of error in the numerical side
are the determinations of the zero momentum mass $\mr$; 
apart from the statistical errors 
one has the systematic errors in the procedures to extract $\mr$
from the (finite volume) data particularly in the symmetry broken phase.   

In the next section we present a summary of the available raw data
in both the symmetric and broken phases. We have 
performed simulations in both phases and in particular increased
the statistics at previously measured $\kappa$ values in the broken 
phase by a factor of $\sim 10$.

We then discuss various determinations of $\Zrhat$ from the data
and compare with theoretical expectations.  
Next we show that these data are not in contradiction
with conventional wisdom. The reason for this conclusion differing
from that of Stevenson has two main origins. Firstly unfortunately our
central value of $\Zrhat$ at $\kappa=0.0751$ in \cite{BDNWW} is 
about one standard deviation lower than that 
obtained from the present run.
In this connection we remark that in those runs we were not 
aiming at high precision but only sufficient to reach our goal 
to present evidence that $\Zrhat$ is not increasing 
logarithmically as one approaches $\kappac$. Secondly we point out that
there is a quantitative uncertainty on the $\rmO(a^2)$
lattice artifacts which are of course increasingly relevant as one goes 
away from the critical point. 
This is of course not at all new, however sometimes
forgotten and perhaps underestimated in standard RG analyses.

\section{Ising MC simulation and results}

We work on hypercubic lattices of volume $L^4$ with periodic
boundary conditions in each direction and with standard action.  
In this paper we adopt the notations and definitions in ref.~\cite{BDNWW},
and will generally not repeat them here. 

If one just wanted to obtain Stevenson's step, measurements
at only two $\kappa$ values are required. However precise simulations
at these points are CPU expensive and hence it is useful to
compute also at points with smaller correlation length to observe the
approach to Stevenson's chosen points which hopefully also reflects the 
approach to the critical point.
As mentioned above we have simulated in both phases. In the symmetric
phase this served as a check with respect to previous simulation results 
by other authors. In Table~\ref{Tab3s} we collect the (to our 
knowledge) best data for observables in the symmetric phase,
which are obtained without a fitting procedure;
by ``best" we mean data with the largest statistics 
and for lattices with large physical volumes $mL\gtrsim6$.
(Our tables can be found in Appendix~B.)
All the values in this table are from the present simulation 
except those for $\kappa=0.07102$ which come from ref.~\cite{MW}.
The previous results of Montvay, M\"{u}nster and Wolff \cite{MMW}
for their lattices A,C ($\kappa=0.071, L=12$
and $\kappa=0.0732, L=20$) are in complete agreement with ours, but
have larger errors. 
\footnote{Some of the entries for their lattice B
($\kappa=0.0724, L=16$) are many standard deviations away from ours
and almost certainly wrong (as also suspected
by Stevenson \cite{Stevenson}).} 
Note that we have also measured at a new point $\kappa=0.07436$ 
slightly closer to $\kappac$ than previous ones. 

Also included in Table~\ref{Tab3s} is the quantity
\be
\mrnaive\equiv
\sqrt{\frac{\widehat{k_0}^2\widetilde{G}(k_0)}
{\widetilde{G}(0)-\widetilde{G}(k_0)}}
\,,\,\,\,\,k_0=\left(\frac{2\pi}{L},0,0,0\right)\,,  
\end{equation}
where $\widetilde{G}$ is the Fourier transform of the (connected)
two--point function. $\mrnaive$ tends to the desired 
zero momentum mass in the infinite volume limit
\footnote{and is volume independent for the free lattice theory with
standard action}.

In Table~\ref{Tab3b} we present a similar data collection for the broken 
phase. This is an update of Table~3 in \cite{BDNWW};
here we have increased our statistics by a factor of $\sim10$.
In these runs we also measured the connected 3--point function $\chi_3$
(except for one lattice).  
For a large subset of the runs in the broken phase we also measured
the correlation matrix of the time slice field $\varphi(t)$
with the composite field 
\be
\varphi_2(t)\propto\varphi(t)^2\,,
\,\,\,\,\varphi(t)=\sum_{\bf x}\phi(t,{\bf x})\,.
\end{equation}

\section{Determination of $\mr$ and derived quantities}

The usual expressions for the wave function renormalization constant 
and renormalized couplings involve the infinite volume zero momentum mass. 
In particular a precise determination of $\Zrhat$ 
(as required to discuss Stevenson's step) requires an equally reliable 
determination of $\mr$. 
Firstly in considering the $r$--point functions one can
verify using the formulae given in ref.~\cite{JMMTW} that 
finite volume effects coming from tunneling are negligible 
for our lattices. We then adopted two fitting procedures. 

In the first we made
fits of $\widetilde{G}(k)^{-1}$ for 
$k=\left(2\pi s/L,0,0,0)\right)$
for small values of $s$ ($s\le3$ in the symmetric phase and $s\le4$ in the 
broken phase) to polynomials in $\widehat{k}^2$
and just recorded good fits with $\chi^2/{\rm dof}<1$. 

Due to the discreteness of available values of $k$ the determination
of the slope at $k=0$ is prone to discretization error.
A removal of this deficit is attempted in a second procedure where we 
performed fits of the connected (time slice) correlation 
function $S(t)$ to obtain the physical mass $m$. 
Relying on the expectation that finite
volume effects on $m$ are extremely small in this model when $mL>7$,
we obtain an estimate for the infinite volume $\mr$ by computing the
second moment with the data in a large portion of the lattice volume
\footnote{e.g. $t\lesssim3/m$ where the errors are reasonably small} 
and then computing the contribution from the rest of the (infinite) volume 
using the measurement of $m$. 

\begin{figure}
\begin{center}
\psfig{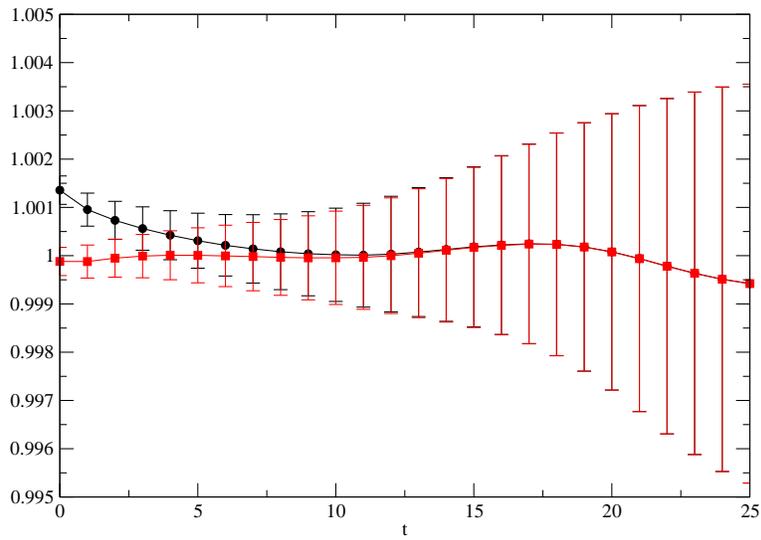}
\end{center}
\caption{\footnotesize Ratio of the measured $S(t)$ to the $1-$ and 
$2-$mass contributions (filled circles and squares respectively)
to the $2-$mass fit for the $\kappa=0.0744, L=52$ lattice.}
\label{fig2}
\end{figure}

\begin{figure}
\begin{center}
\psfig{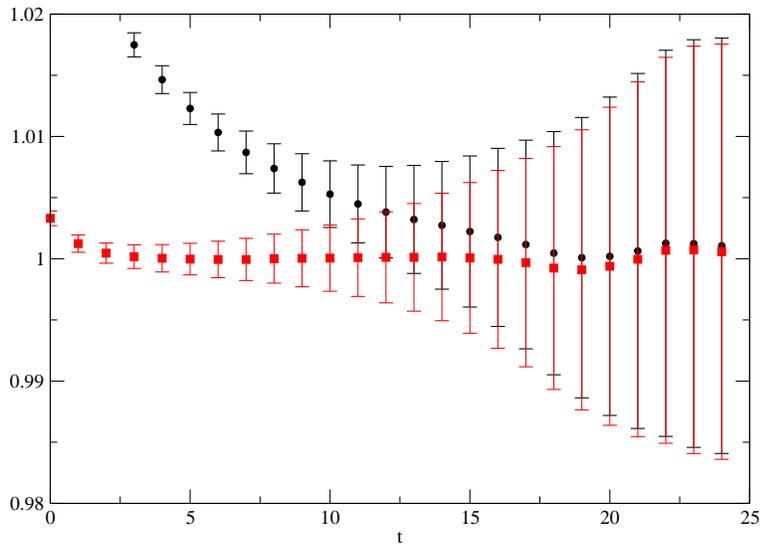}
\end{center}
\caption{\footnotesize As for Fig.~\ref{fig2} but  
for the $\kappa=0.0751, L=48$ lattice.}
\label{fig3}
\end{figure}

Our fits of $S(t)$ were in terms of one and 
two mass cosh functions; for the $1-$mass case we fitted only distances
$t>1/m$ and for the $2-$mass case $t\gtrsim 2/3m$. The $2-$mass fits were
constrained to have the second mass fixed to $3m$ for the symmetric
phase and $2m$ for the broken phase. The $2-$mass fits give (as expected)
a slightly lower central value of $m$ but a slightly larger error
than the $1-$mass fit. This is again reflected in the resulting values 
for $\mr$ and $\Zrhat$. Figs.~\ref{fig2},~\ref{fig3} are typical
examples in the symmetric and broken phases respectively, 
which illustrate the good quality of the fits and the relatively 
very small contribution of the higher particle states. 

For the subset of data in the broken phase where we had the full 
correlation matrix 
mentioned above we found that the two operators $\varphi,\varphi_2$ were 
nearly parallel, both coupling very
weakly to the 2--particle state, and thus this did not help to 
significantly reduce the errors on $m$. 

In Tables~\ref{Tab4s} and \ref{Tab4b} we give results for $m$  and  
quantities derived from various
estimates of $\mr$ in the symmetric and broken phase respectively.
In practically all cases all methods to determine $\mr$ gave 
compatible results 
\footnote{One exception is the lattice $\kappa=0.0752$ where
a best fit gave a slightly higher value of $\mr$. The resulting
value of $\Zrhat$ is bigger than that at $\kappa=0.0751$ which 
we consider as a signal of
the potential instability of such momentum space fits.}.

\begin{figure}
\begin{center}
\psfig{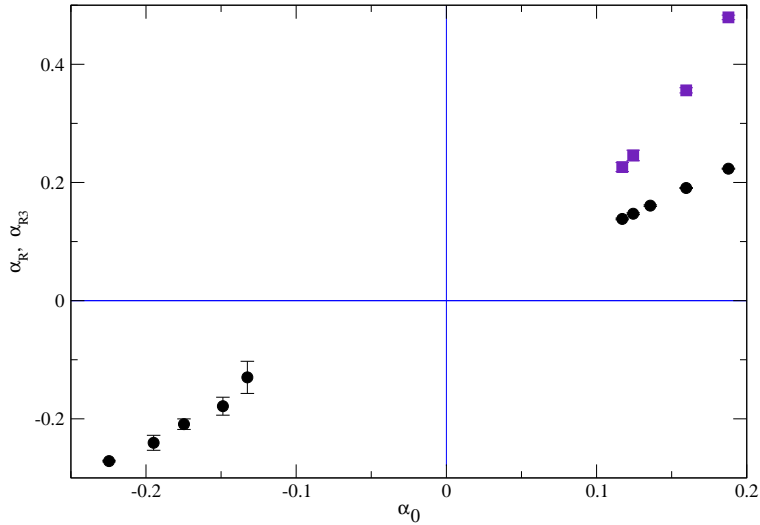}
\end{center}
\caption{\footnotesize Renormalized couplings versus $\alpha_0$;
$\alpha_0<0$ corresponds to the symmetric phase and $\alpha_0>0$
to the broken phase. Full circles and squares represent $\alphar$
and ${\alphar}_3$, respectively. For $\alpha_0<0$ we have plotted
$-\alphar$.}
\label{fig4}
\end{figure}

In Fig.~\ref{fig4} we plot 
renormalized couplings ($\alphar=\gr/(16\pi^2)$) in the  
symmetric and broken phases using $\mr=\mrnaive$ versus 
\be
\alpha_0(\kappa)\equiv 
\frac{2\,{\rm sign}(\tau)}{3\ln|\tau|}\,,
\,\,\,\,\,\,\,\,\tau=1-\frac{\kappa}{\kappac}\,,
\end{equation}
with the presently best estimate of $\kappac=0.074848$ from 
ref.~\cite{StaufAd}.
In the symmetric phase the coupling is defined through the 
connected 4--point function. In the broken phase we have included the
renormalized coupling $\gr$ defined through the vacuum expectation value
and another coupling $\grthree$ defined through the connected 3--point 
function. According to standard RG analysis these should behave as 
$16\pi^2\vert\alpha_0\vert$ as $\tau\to0$. Note that the relations of
the measured values for $\gr$ and $\grthree$ are quite consistent with 
those obtained in renormalized perturbation theory:
\be
\grthree = \gr\left(1+3\alphar+3.75\alphar^2+\dots\right)\,.
\end{equation}

\subsection{Stevenson's step}

\begin{figure}
\begin{center}
\psfig{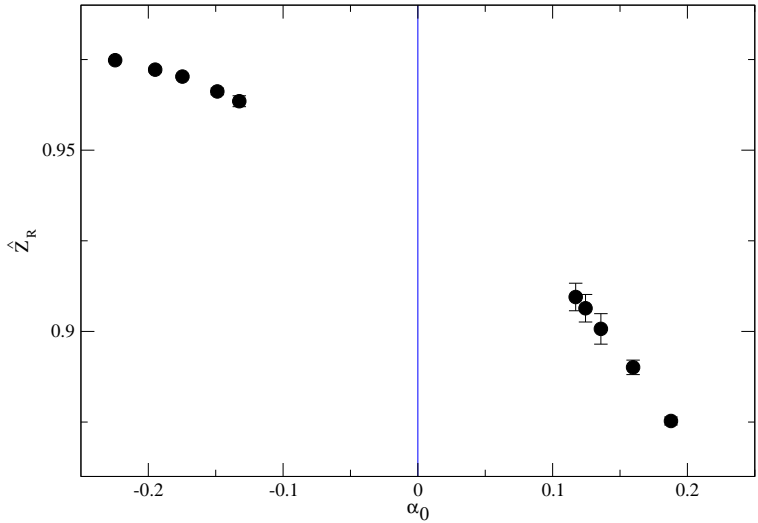}
\end{center}
\caption{\footnotesize Measured values of $\Zrhat$ using $\mr$ obtained
by the 1-mass fit method.}
\label{fig5}
\end{figure}

\begin{figure}[htb]
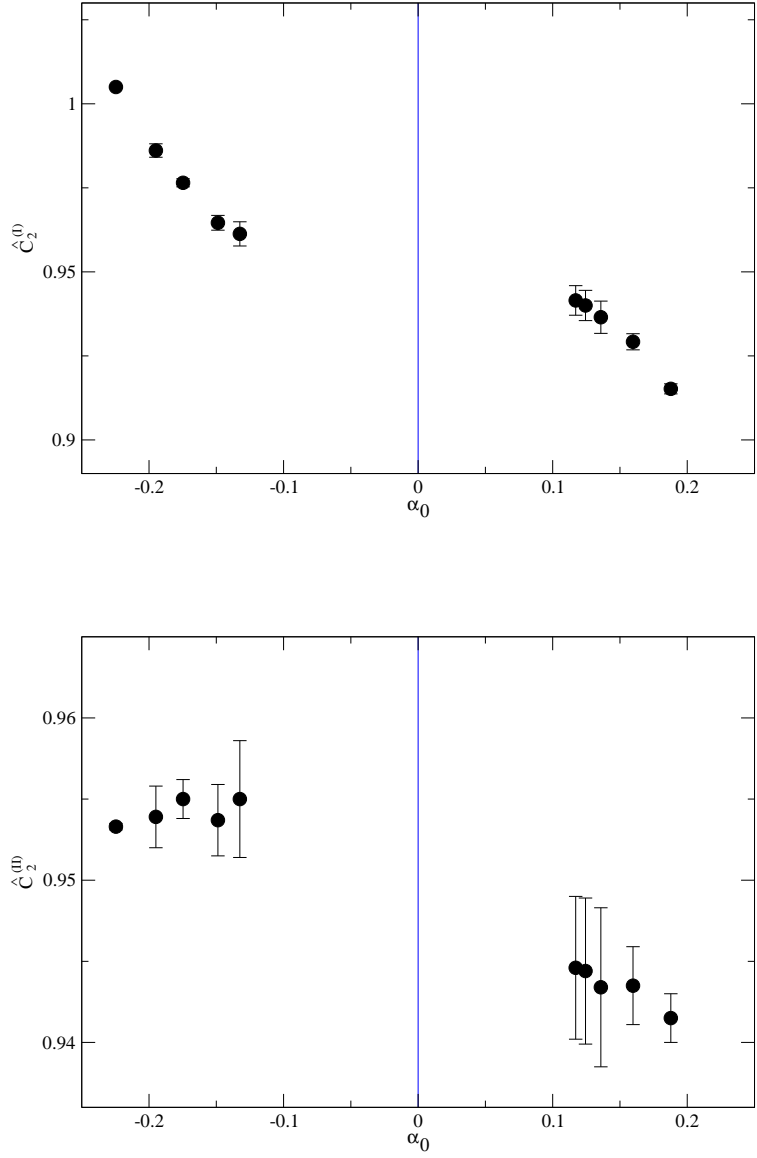

\begin{center}
\epsfxsize=0.71\linewidth
\epsfbox{C2_I1.eps}
\vskip 15mm 
\epsfxsize=0.71\linewidth
\epsfbox{C2_II1.eps}
\end{center}
\caption{{}\footnotesize
Top, $C_2^{(I)}$ (renormalization scheme of refs. [2,3]). 
Bottom, $C_2^{(II)}$ (field theoretical renormalization scheme). 
Note the different vertical scales.}
\label{fig6}
\end{figure}

Stevenson presents plots with respect to $\kappa$ 
whereas we prefer to present plots wrt $\alpha_0$ 
since dependence of quantities of interest
on this variable are expected to be smoother.
In Fig.~\ref{fig5} we plot $\Zrhat$ obtained using $\mr$ values
obtained from the 1-mass fit method. Using $\mr=\mrnaive$ or $\mr$
corresponding to the 2-mass fit method would result in very similar
plots, with somewhat larger errors for the latter case. (The same
applies also to Fig.~\ref{fig6}.) 
Here one certainly
does not see any signal of a discontinuity at the critical point.
As mentioned above the value of $\Zrhat(\kappa=0.0751)$ cited in 
\cite{BDNWW} is over one standard deviation away from our present value
\footnote{Also note that our present value of $\Zrhat$ at $\kappa=0.0754$ 
has moved down wrt to that quoted in \cite{BDNWW}, and is now consistent
with some as yet unpublished data by Cea and Cosmai mentioned in 
\cite{Stevenson}}. 
For Stevenson's step we would now quote a measured value of
\be
\triangle_{\rm MC} = 0.057(6)\,.
\end{equation}
Given the more precise value of $\Zrhat$ at $\kappa=0.0751$
Stevenson may not have written his paper. On the other hand
this new value for $\triangle$ is still bigger than what Stevenson 
calls analytically feasible. 
We however disagree with this opinion. 
The problem is that our quantitative control of $\rmO(a^2)$ 
effects in the RG equations is not sufficient. Although these are rather
small at the values of $\kappa$ used in the definition of $\triangle$,
they are still not negligible when discussing possible discrepancies 
between theory and ``experiment". In refs.~\cite{LWsymm,LWbroken} 
some such effects 
were taken into account by including the $\mr$ dependence
of perturbative RG coefficients appearing at low orders of perturbation 
theory. This procedure was also adopted by Stevenson \cite{Stevenson}. 
It is however just a pragmatic procedure 
(i.e. practically the best one could quantitatively
do at the time), but it is not a quantitatively systematic prescription. 
Firstly it is not consistent to include $\rmO(a^2)$
effects while ignoring higher perturbative effects
\footnote{analogous procedures are often (similarly questionably) adopted 
in phenomenology when taking higher twist effects into account.}. 
Secondly even if one disregards this, the leading $\rmO(\mr^2)$ effects
at $r$--loop order are of generically of the form $\gr^{c+r}\mr^2|\ln\mr|^r$ 
and hence all quantitatively of the same order  
since the RG equations predict $\gr\sim 1/\ln\mr$.

To obtain the leading cutoff effects one
must follow the method of Symanzik~\cite{Sym}. The analysis shows
that the leading artifacts 
for the correlation functions (for the case $n=1$) are of the form 
\be
\Gamma\sim\Gamma^{(0)}+\rmO\left(\mr^2|\ln\mr|^{1/3}\right).
\label{Gamma}
\end{equation}
Here $\Gamma^{(0)}$ is the formal perturbative sum; in the process of
defining this non-perturbatively it could be that
renormalon-type effects would lead to cutoff effects of the same form 
as the leading operator insertion. But even though the form of the
leading cutoff corrections may be known 
the amplitude is undetermined. 

Another simple exercise to appreciate this point is to 
compute in the leading order of the $1/n$ expansion. In that
limit the cutoff effects are of the order $\mr^2$ however these
are not given by taking the limit $n\to\infty$ of the first perturbative
contributions. Some illustrations are given in Appendix~A.  

Having realized that we have unfortunately insufficient quantitative
knowledge of $\mr^2$--effects it is still legitimate to
ask if the numerically found value of $\triangle$ looks inconsistent
with the conventional theoretical expectation that $\Zrhat$ approaches the 
same value $\widehat{C}_2$ coming from both sides of the critical point:
\be
{\rm CW}:\,\,\,\,\,
\lim_{\kappa\to\kappac^+}\Zrhat=\widehat{C}_2
=\lim_{\kappa\to\kappac^-}\Zrhat\,,\,\,\,\,\widehat{C}_2=2\kappac C_2\,. 
\end{equation}
To illustrate the situation we consider two 
expressions $\widehat{C}_2^{(I)}$ and $\widehat{C}_2^{(II)}$ which
we would eventually expect to approach $\widehat{C}_2$ faster than 
$\Zrhat$. The first, which arises in the renormalization scheme of 
\cite{LWsymm} is defined by dividing $\Zrhat$ by its 
perturbative expansion truncated at 2--loops:
\be
\widehat{C}_2^{(I)}\equiv
\left\{
\begin{split}
&\Zrhat 
\left(1+\frac{1}{18}\alphar+0.100896\alphar^2\right)^{-1}\,,
\,\,\,\,\,\,{\rm for}\,\,\kappa<\kappac\,,
\\
&\Zrhat
\left(1-\frac{7}{36}\alphar-0.538874\alphar^2\right)^{-1}\,,
\,\,\,\,\,\,{\rm for}\,\,\kappa>\kappac\,.\\
\end{split}
\right.
\end{equation}
The second, which is a natural choice in a field theoretical context,
is merely defined from the first by 
\be
\widehat{C}_2^{(II)}\equiv\frac{\kappa}{\kappac}\widehat{C}_2^{(I)}\,.
\end{equation}
The difference between the two functions is just an order $\mr^2$ cutoff
effect \cite{LWsymm,LWbroken}:
\ba
\kappac-\kappa&=&\phantom{\frac12}
C_3\mr^2\gr^{-1/3}\left\{1+\rmO(\gr)\right\}\,,
\,\,\,\,{\rm for}\,\,\kappa<\kappac\,,
\\
\kappa-\kappac&=&\frac12 C'_3\mr^2\gr^{-1/3}\left\{1+\rmO(\gr)\right\}\,,
\,\,\,\,{\rm for}\,\,\kappa>\kappac\,,
\ea
which incidentally has the same cutoff effects as in (\ref{Gamma}).
In Fig.~\ref{fig6} we plot them together to give an idea on the 
importance of the $\rmO(\mr^2)$ effects. Both are certainly 
not inconsistent with the expectation that the limits are the same
on both sides (note the difference in scales on the vertical axis). 
Indeed if one allowed to naively

\clearpage

\noindent
extrapolate the 
curves by eye one would probably infer a different sign for the
difference of the limiting values from the two plots.

If one had much more precise data one could attempt constrained
fits to the CW but this is not warranted with the present data.
Assuming CW we would now quote $\widehat{C}_2=0.95(1)$ which
corresponds to $\ln C_2=1.85(1)$ 
to be compared with the result $\ln C_2=1.87(1)$ in 
\cite{LWbroken}.


\section{Reply to some other critiques}

In ref.~\cite{CCC2} the authors reconsider the quantity 
$v^2\chi$ in the broken phase which according to CW behaves as
\be
v^2\chi=a_1\left(\ell-\frac{25}{27}\ln\ell\right)+a_2+\dots\,,
\label{v2chi}
\end{equation}
where $\ell\equiv|\ln(\kappa-\kappac)|$ with
\ba
a_1&=&\frac{9C_2^2}{32\pi^2}\,,
\\
\frac{a_2}{a_1}&=&\ln C'_3+2\ln C'_1-1.6317\,.
\ea
A fit of the expression (\ref{v2chi}) to the data gave \cite{BDNWW}
$a_1=1.267(14)$ and $a_2=-2.89(8)$ whereas the theoretical prediction
based on the results quoted in \cite{LWbroken} for the values of the 
non-perturbative constants $C_1$, $C_2$ and $C_3$ is $a_1=1.20(3)$ and
$a_2=-1.6(5)$. The authors of ~\cite{CCC2} claim that such a comparison
``shows that the quality of the 2-loop fit is poor". How they can reach
such a conclusion is surprising to us. Firstly the values $a_1$ agree
within one standard deviation. Secondly the value of $a_2$ from the
fits can only be regarded as effective. Higher order terms 
e.g. of the form $1/\ell$ are completely neglected in the fit.

\begin{figure}
\begin{center}
\psfig{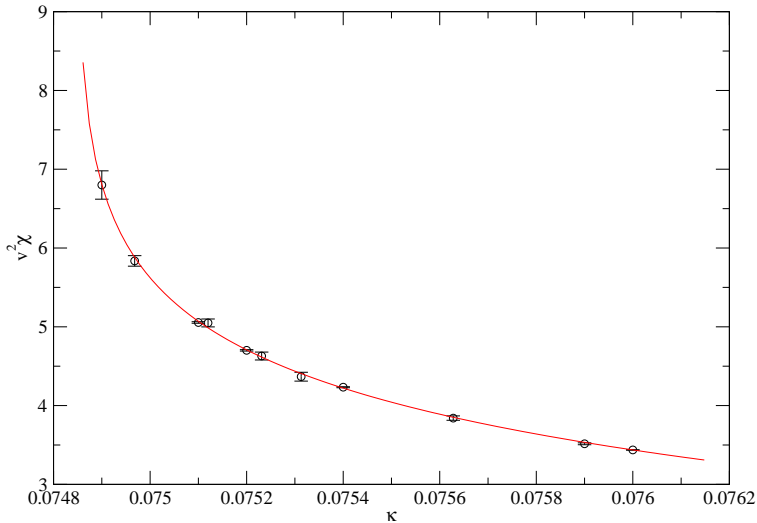}
\end{center}
\caption{\footnotesize 2-parameter fit of the expression (\ref{v2chi})
to the data from Table~\ref{Tab3b}.} 
\label{figfit}
\end{figure}

We have made new fits including all data from Table~\ref{Tab3b}, omitting 
only the $\kappa>0.076$ lattices and the one corresponding to
$\kappa=0.07504$ (because of its too small physical volume). Our first
fit, which is shown in Fig.~\ref{figfit}, has an acceptable 
$\chi^2/{\rm dof}=1.2$ and yields 
\begin{equation}
a_1=1.224(6)\qquad\qquad{\rm and}\qquad\qquad a_2=-2.68(3).
\label{fit1}
\end{equation}
As discussed at the end of the previous section, our $\Zrhat$ 
data prefer a slightly smaller $C_2$ value corresponding to 
$\ln C_2=1.85(1)$. This changes the theoretical prediction to $a_1=1.15(3)$.
We have made a second two-parameter fit with a third term of the form
$a_3/\ell$ added but where the first coefficient was kept fixed at 
$a_1=1.15$. The result of this fit\footnote{which has incidentally
a value of $a_2$ quite close to the prediction of
ref. \cite{LWbroken}} is 
\begin{equation}
a_1=1.15,\qquad a_2=-1.80(3),\qquad a_3=-3.5(2),\qquad \chi^2/{\rm dof}=1.0.
\label{fit2}
\end{equation}
The fact that it is impossible to distinguish between (\ref{fit1}) and 
(\ref{fit2}) illustrates the point we made in the previous paragraph.
Taking into account all the uncertainties we find the agreement between
theory and MC \lq experiment' satisfactory.

\begin{figure}
\begin{center}
\psfig{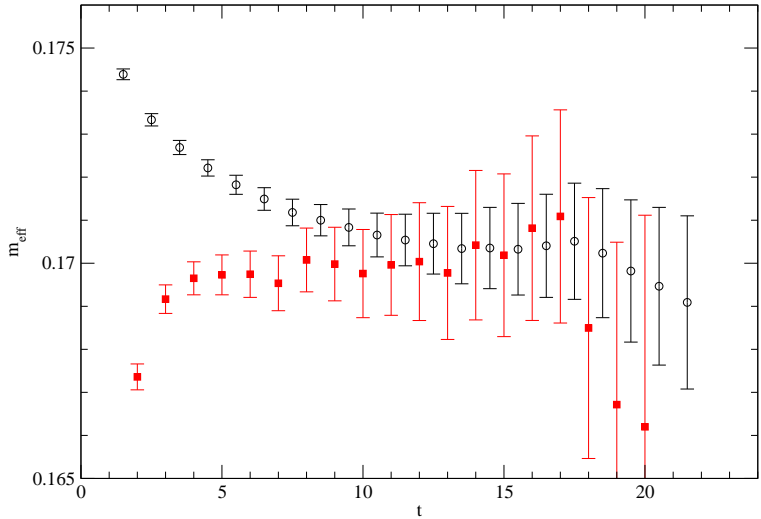}
\end{center}
\caption{\footnotesize Effective masses for the $\kappa=0.0751, L=48$ 
lattice. The circles are from the 1-mass definition, and the squares
from the 2-mass fit definition as described in the text.}
\label{fig1}
\end{figure}

We completely disagree with the implications of Section 5 in 
ref.~\cite{Stevenson}. Here Stevenson proposes that a proper extraction
of $\mr$ is obtained by globally fitting the data for 
$\widetilde{G}(k)^{-1}$ in the whole available 
momentum range to the form of the free lattice propagator with standard 
(nearest neighbor) action.
But this is an a priori incorrect procedure since $\mr$ must be extracted 
from data including only low momenta. 
Accepting this fact it is then a rather surprising ``experimental" finding 
that the numerically measured propagator is so extremely close to the 
naive propagator; only a detailed look reveals that there is 
some significant deviation at larger momenta. It is a bit easier
to see the deviation in coordinate space e.g. in
Fig.~\ref{fig1} we plot the effective masses at $\kappa=0.0751$.
The effective mass $m_{\mathrm{eff}}(t+0.5)$ is defined for the 1--mass
case by $m_1$ in the ansatz $B\left[\exp(-m_1t')+\exp(-m_1(L-t'))\right]$
with the parameters defined from $S(t')$ at  $t'=t,t+1$. This
would be constant for a free standard lattice propagator, but the data
shows clear $t$-dependence.  
For the constrained 2--mass fit 
$\sum_{i=1}^2 B_i\left[\exp(-m_it')+\exp(-m_i(L-t'))\right]$
with $m_2=2m_1$, $m_{\mathrm{eff}}(t)=m_1$ from the correlation function
at $t'=t-1,t,t+1$.

\section{Conclusions}

An experimental observation contradicting a prediction of an
until then accepted theory is always an exciting event.
It invalidates the theory as it stands
and inevitably leads to progress in our understanding. 
Similarly finding mismatches between theoretical predictions 
and numerical simulations in the $\phi^4_4$ theory as claimed 
in refs.~\cite{CCC2,Stevenson} 
would be a serious blow if they withheld scrutiny.
We hope to have convinced the reader in this paper that 
conventional wisdom concerning this structurally simple theory 
is still alive. Although present numerical simulations support CW, 
the scenario can unfortunately only be ``nailed down" by analytic proofs.  

\subsection{Acknowledgments}

We thank the Leibniz-Rechenzentrum where part of the computations
were carried out.
This investigation was supported in part by the Hungarian 
National Science Fund OTKA (under T049495 and T043159)
and by the Schweizerischer Nationalfonds.

\vfill
\eject

\clearpage
\newpage
\appendix
\renewcommand{\thesection}{Appendix~A: leading order $1/n$ expansion}
\section{}
\renewcommand{\thesection}{A}

For the $n$-component model we take over the notations of ref.~\cite{LWO4}.
In the $1/n$ expansion one takes $n\to\infty$ with
\be
\lambdahat\equiv n\lambda
\end{equation}
held fixed.

In the symmetric phase in leading order we have
\be
\Zrhat=1\,,
\end{equation}
and given $\lambdahat,\kappa<\kappa_{\rm c}=\frac12 J_1(0)$ 
the renormalized zero momentum mass $\mr$ is determined by 
\be
\frac{\kappa^2}{\lambdahat}\left(\mr^2+8-\frac{1}{\kappa}\right)
+2\kappa-J_1(\mr)=0\,,
\end{equation}
where (see ref.~\cite{LWsymm})
\be
J_q(\mr)=\int_{-\pi}^\pi \frac{\rmd^4 k}{(2\pi)^4}
\left(\hat{k}^2+\mr^2\right)^{-q}\,.
\label{B4}
\end{equation}

The renormalized coupling $\grhat\equiv n\gr$ is given by
\be
\grhat=\frac{6}{\frac{\kappa^2}{\lambdahat}+J_2(\mr)}\,.
\end{equation}
For $\lambdahat=\infty$ there is a simplification and
\ba
\mr\frac{\partial\grhat}{\partial\mr}
&=&\frac{24\mr^2 J_3(\mr)}{J_2(\mr)^2}
\\
&=&\frac{s_1}{3}\grhat^2+\mr^2 B_1(\mr)\,,
\ea
with $s_1=1/(16\pi^2)$ and
\be 
B_1(\mr)=\frac{24\left(J_3(\mr)-\frac{s_1}{2\mr^2}\right)}{J_2(\mr)^2}\,.
\end{equation}
With the expansions of $J_q(\mr)$ in \cite{LWsymm} 
for small $\mr$ the non-scaling piece behaves as
\be
B_1(\mr)\sim -\frac{3}{s_1\ln(\mr^2)}+\dots
\label{lnonscaling}
\end{equation}

In renormalized PT one obtains (in the limit $n\to\infty$):
\be
\mr\frac{\partial\grhat}{\partial\mr}
=\frac{4\mr^2}{8+\mr^2}\grhat
+\grhat^2\left(\frac{s_1}{3}+\frac{2\mr^2}{3}
\left[J_3(\mr)-\frac{J_1(\mr)}{(8+\mr^2)^2}\right]\right)
+\rmO(\grhat^3\mr^2)\,.
\end{equation}
So including the leading non-scaling effects we write
\be
\mr\frac{\partial\grhat}{\partial\mr}
=\frac{s_1}{3}\grhat^2+\frac{\mr^2}{\ln(\mr^2)}\sum_{s=0}f_s
+\rmO(\mr^2/\ln^2\mr^2)
\end{equation}
where $f_s$ denotes the coefficient from the $s$ loops. For the first
coefficients we have
\be
f_0=-\frac{3}{s_1}\,,\,\,\,\,
f_1=-\frac{3}{s_1}\,,
\end{equation}
to be compared to the non-perturbative result in Eq.~(\ref{lnonscaling}).

Although $\Zrhat$ is so simple, the $\gamma$--function defined in 
\cite{LWO4} is not zero e.g. in the limit $\lambdahat\to\infty$:
\be
v \equiv\frac12\mr\frac{\partial}{\partial\mr}\ln\Zr
=\mr^2 \frac{J_2(\mr)}{J_1(\mr)}\,,\,\,\,\,\,\,\,\,\,(\lambdahat=\infty)\,.
\end{equation}
For small $\mr$ we have
\be
v \sim -\frac{s_1}{J_1(0)}\mr^2\ln(\mr^2)+\dots
\end{equation}
whereas in renormalized perturbation theory one obtains in the leading
order $1/n$ expansion
\ba
v &=&\frac{\mr^2}{8+\mr^2}\left[1
+\frac16\left(J_2(\mr)-\frac{J_1(\mr)}{8+\mr^2}\right)\grhat
+\rmO(\grhat^2)\right]
\\
&\sim& \frac{\mr^2}{8}\left[1+1+\rmO(\grhat^2)\right]\,,
\ea
which has no $\mr^2\ln(\mr)$ behavior at tree and 1-loop level
whereas the full non-perturbative function does.

Similar features are found in the symmetry broken phase.

\vfill
\eject

\newpage
\appendix
\renewcommand{\thesection}{Appendix~B: Tables}
\section{}
\renewcommand{\thesection}{B}

\vspace{5.0cm}

\begin{table}[ht]
\centering
\begin{tabular}[t]{l|l|l|l|l|l}
\hline
$\,\,\,\,\,\kappa$&$\,L$&$\,\,\,\,\,\,\,\,\,\,\,\chi$
&$\,\,\,\,\,\,\,\,\,-\chi_4$&$\,\,\mrnaive$
\\[1.0ex]
\hline \hline
$0.071$   &$12$&$27.9884(29)$ &$5.602(26)\cdot 10^5$
&$0.49528(4)$  \\[1.0ex]
$0.071$   &$18$&$28.0184(12)$ &$5.606(26)\cdot 10^5$
&$0.49500(2)$  \\[1.0ex]
$0.07102$   &$12$&$28.20(3)$ &$5.9(6)\cdot 10^5$
&$0.4924(5)^*$\\[1.0ex]
$0.0724$   &$24$&$45.697(14)$ &$3.68(19)\cdot 10^6$
&$0.38330(14)$  \\[1.0ex]
$0.0732$   &$20$&$69.991(21)$ &$1.749(44)\cdot 10^7$
&$0.30760(8)$  \\[1.0ex]
$0.0732$   &$30$&$70.009(9)$ &$1.768(50)\cdot 10^7$
&$0.30764(4)$  \\[1.0ex]
$0.074$   &$40$&$142.676(62)$ &$2.74(23)\cdot 10^8$
&$0.21392(9)$  \\[1.0ex]
$0.07436$   &$52$&$257.27(19)$ &$2.14(45)\cdot 10^9$
&$0.15871(12)$  \\[1.0ex]
\hline
\end{tabular}
\caption{\footnotesize Measured values of $\chi,\chi_4,$ and 
$\mrnaive$ in the symmetric phase of the Ising model.
The star indicates that for this entry 
$\mrnaive$ was not measured directly, but could be estimated from
the published data.}
\label{Tab3s}
\end{table}

\begin{table}[ht]
\centering
\begin{tabular}[t]{l|c|c|l|l|l|c}
\hline
$\,\,\,\,\,\kappa$&$L$&$\chi$&$\,\,\,\,\,\,\,\,v$
&$\,\,\,\mrnaive$& \ \ \ \ $-\chi_3$& ref.\\[1.0ex]
\hline \hline
$\,\,\,\kappa_0$&$10$&$5.130(2)$&$0.571267(8)$
&$1.0041(5)^+$&&\cite{BDNWW}\\[1.0ex]
\hline
$0.077$&$16$&$18.196(9)$&$0.389500(9)$&$0.5595(5)$&$3.734(22)\cdot 10^3$ &
\\[1.0ex]
$0.077$&$32$&$18.21(4)$&$0.38951(1)$&&&\cite{CCC}\\[1.0ex]
\hline
$0.076$&$20$&$37.80(3)$&$0.301544(13)$&$0.3942(5)$&$1.942(14)\cdot 10^4$ &
 \\[1.0ex]
$0.076$&$32$&$37.70(31)$&$0.3015(1)$&$0.428(5)^*$&&\cite{CCCS}\\[1.0ex]
\hline
$0.0759$&$32$&$41.71(13)$&$0.29030(2)$&&&\cite{CCC}\\[1.0ex]
$0.0759$&$48$&$41.95(93)$&$0.29028(5)$&&&\cite{CCC}\\[1.0ex]
\hline
$0.075628$&$48$&$58.70(42)$&$0.25580(2)$&&&\cite{CCC}\\[1.0ex]
\hline
$0.0754$&$32$&$87.08(13)$&$0.220488(14)$&$0.2622(6)$&\\[1.0ex]
\hline
$0.075313$&$48$&$104.2(1.3)$&$0.20477(4)$&&&\cite{CCC}\\[1.0ex]
\hline
$0.075231$&$60$&$130.8(1.4)$&$0.18812(3)$&&&\cite{CCC}\\[1.0ex]
\hline
$0.0752$&$36$&$142.1(8)$&$0.18138(5)$&$0.2054(16)^+$&&\cite{BDNWW}\\[1.0ex]
$0.0752$&$40$&$143.03(28)$&$0.181291(14)$&$0.2054(6)$&$ 4.374(80)\cdot
10^5$& \\[1.0ex]
$0.0752$&$48$&$142.6(9)$&$0.18132(4)$&$0.2055(18)^+$&&\cite{BDNWW}\\[1.0ex]
\hline
$0.07512$&$32$&$193.1(1.7)$&$0.1617(1)$&$0.206(4)^*$&&\cite{CCCS}\\[1.0ex]
\hline
$0.0751$&$48$&$206.32(40)$&$0.156532(12)$&$0.1715(5)$&$1.043(19)\cdot
10^6 $&
\\[1.0ex]
$0.0751$&$52$&$201.2(6.2)$&$0.15654(7)$&&&\cite{CCC}\\[1.0ex]
$0.0751$&$60$&$202.4(8.6)$&$0.15648(2)$&&&\cite{CCC}\\[1.0ex]
\hline
$0.07504$&$32$&$293.4(2.9)$&$0.13822(12)$&$0.172(3)^*$&&\cite{CCCS}\\[1.0ex]
\hline
$0.074968$&$68$&$460.2(4.9)$&$0.11261(5)$&&&\cite{CCC}\\[1.0ex]
\hline
$0.0749$&$68$&$1125(36)$&$0.07736(12)$&&&\cite{CCC}\\[1.0ex]
$0.0749$&$72$&$1141(39)$&$0.07752(21)$&&&\cite{CCC}\\[1.0ex]
\hline
\end{tabular}
\caption{\footnotesize
Measured values of $\chi,v$, $\mrnaive$ and $\chi_3$
from various Ising simulations; data from this investigation have no entry 
in the last column. $\kappa_0=0.080795$. 
In the $\mr$ column $^+$ indicates that instead of the naive, the 1-mass
fit method has been used and $^*$ indicates the $m_{latt}$ mass of 
ref.~[16].}
\label{Tab3b}
\end{table}

\begin{table}[ht]
\centering
\begin{tabular}[t]{l|l|l|l|l|l|l}
\hline
$\,\,\,\,\,\kappa$&$\,\,L$&$\,\,{\rm iter}$&$\,\,\,\,\,\,\,\,\,\,\,m$
&$\,\,\,\,\,\,\,\,\,\mr$
&$\,\,\,\,\,\,\Zrhat$&$\,\,\,\,\,\gr$\\[1.0ex]
\hline \hline
$0.071$   &$18$&$34M$ &$0.49007(2)$&$0.49499(2)$  &$0.9748(1)$ 
&$42.87(20)$\\
&&&&$0.49500(2)$&$0.9748(1)$\\
&&&&$0.49499(2)$&\\
\hline
$0.0724$   &$24$&$2.1M$ &$0.38100(18)$&$0.38331(14)$  &$0.9722(7)$ 
&$38.0(2.0)$\\
&&&&$0.38330(14)$&$0.9722(5)$\\
&&&&$0.38330(16)$&\\
\hline
$0.0732$   &$30$&$10.1M$ &$0.30645(5)$&$0.30765(5)$  &$0.9701(3)$ 
&$32.3(9)$\\
&&&&$0.30764(4)$&$0.9700(2)$\\
&&&&$0.30757(4)$&\\
\hline
$0.074$   &$40$&$1.4M$ &$0.21347(12)$&$0.21391(10)$  &$0.9662(9)$ 
&$28.2(2.4)$\\
&&&&$0.21392(9)$&$0.9663(5)$\\
&&&&$0.21395(8)$&\\
\hline
$0.07436$   &$52$&$615k$ &$0.15850(16)$&$0.15869(13)$  &$0.9635(15)$ 
&$20.5(4.3)$\\
&&&&$0.15871(12)$&$0.9638(9)$\\
&&&&$0.15870(9)$&\\
\hline
\end{tabular}
\caption{\footnotesize Extracted values of $m,\mr,\Zrhat$ and $\gr$ 
for the Ising model in the symmetric phase.
The various values for $\mr$ and $\Zrhat$ are given using the $x$-space 
1--mass fit, naive and momentum space fit methods (in this order).
\lq iter' is the number of sweeps as  described in Sect.~4.1 of ref.~[11].}  
\label{Tab4s}
\end{table}

\begin{table}[ht]
\centering
\begin{tabular}[t]{l|l|l|l|l|l|l|l}
\hline
$\,\,\,\,\,\kappa$&$\,\,L$&$\,\,{\rm iter}$&$\,\,\,\,\,\,\,\,\,\,\,m$
&$\,\,\,\,\,\,\,\,\,\mr$
&$\,\,\,\,\,\,\Zrhat$&$\,\,\,\,\,\gr$&$\,\,\,\,\,\,\,\, {\gr}_3$\\[1.0ex]
\hline \hline
$0.077$   &$16$&$31M$ &$0.5506(6)$&$0.5589(5)$  &$0.8753(11)$ 
&&\\
&&&$0.5502(8)$&$0.5587(6)$  &$0.8749(13)$ 
&&\\
&&&&$0.5595(5)$&$0.8772(12)$&$35.26(12)$&$75.7(6)$\\
&&&&$0.5593(4)$&$0.8764(10)$&&\\
\hline
$0.076$   &$20$&$14.4M$ &$0.3902(7)$&$0.3936(6)$  &$0.8901(20)$ 
&&\\
&&&$0.3890(12)$&$0.3933(7)$  &$0.8889(25)$ 
&&\\
&&&&$0.3942(5)$&$0.8927(17)$&$30.10(13)$&$56.2(7)$\\
&&&&$0.3940(4)$&$0.8917(13)$&&\\
\hline
$0.0754$   &$32$&$6.2M$ &$0.2606(7)$&$0.2619(6)$  &$0.9007(42)$ 
&&\\
&&&$0.2594(14)$&$0.2616(8)$  &$0.8988(79)$ 
&&\\
&&&&$0.2622(6)$&$0.9028(31)$&$25.40(20)$&\\
&&&&$0.2621(4)$&$0.9020(20)$&&\\
\hline
$0.0752$   &$40$&$4.9M$ &$0.2043(7)$&$0.2053(6)$  &$0.9064(38)$ 
&&\\
&&&$0.2030(13)$&$0.2049(8)$  &$0.9027(56)$ 
&&\\
&&&&$0.2054(6)$&$0.9074(39)$&$23.23(24)$&$38.8(1.4)$\\
&&&&$0.2076(4)$&$0.9192(16)$&&\\
\hline
$0.0751$   &$48$&$6.3M$ &$0.1707(6)$&$0.1713(5)$  &$0.9095(38)$ 
&&\\
&&&$0.1699(11)$&$0.1711(6)$  &$0.9073(53)$ 
&&\\
&&&&$0.1715(5)$&$0.9112(39)$&$21.84(22)$&$35.7(1.2)$\\
&&&&$0.1715(4)$&$0.9113(29)$&&\\
\hline
\end{tabular}
\caption{\footnotesize Measured values of $m,\mr,\Zrhat$, $\gr$ and
${{\gr}}_3$ for the Ising model in the broken phase.
Values for $m,\,\mr$ and $\Zrhat$ are given using the $x$-space 1-mass fit,
2-mass fit, naive and momentum space fit methods (in this order).
\lq iter' is the number of sweeps as  described in Sect.~4.1 of ref.~[11].}  
\label{Tab4b}
\end{table}

\clearpage

\eject

\end{document}